\documentclass[aps, prb, showpacs, twocolumn, amsmath, letterpaper]{revtex4-2}

\usepackage{graphics}
\usepackage{booktabs}

\usepackage{amssymb}
\usepackage{bm}
\setcitestyle{super}
\usepackage{verbatim}
\usepackage[english]{babel}

\usepackage{amsmath}
\usepackage{graphicx}
\usepackage{wrapfig}
\usepackage[colorlinks=true, allcolors=blue]{hyperref}

\begin{document}

\title{Structure-Driven Prediction of Magnetic Order in Uranium Compounds}

\author{Christopher Broyles}
\thanks{Equal contribution}
\author{William Charles}
\thanks{Equal contribution}
\author{Sheng Ran}
\affiliation{Department of Physics, Washington University in St. Louis, St. Louis, MO 63130, USA}

\begin{abstract}

The advancement of machine learning technologies has revolutionized the search and optimization of material properties. These algorithms often rely on theoretical calculations, such as density functional theory (DFT), for data inputs and validation, which are not always effective for uranium-based materials due to their strong electron correlations. This study presents a computationally inexpensive machine learning approach, specifically a random forest classifier, to predict the magnetic ground states of uranium compounds using only structural inputs. Our model, trained on a curated dataset of experimentally-verified magnetic orders, achieves a mean accuracy of 60.2\%, significantly outperforming random chance. By excluding computationally intensive DFT calculations, our method offers a faster and reliable alternative for discovering new materials with desirable magnetic properties, addressing the challenges posed by strong correlations in quantum materials.

%Compared to a DFT-assisted prediction of uranium ternary compounds, our model's classification accuracy is competitive, falling short by only 6\%. 

\end{abstract}

\maketitle{}
\section{Introduction} \label{sec:intro}

Predicting the magnetic order of strongly correlated electron systems is challenging due to their complex nature. Traditional computational methods, such as Density Functional Theory (DFT) with an added Hubbard $U$ correction (DFT+$U$), are often employed to account for electron-electron interactions~\cite{Cococcioni2005,Himmetoglu2014}. However, DFT+$U$ typically does not work well for Kondo lattice systems, where the magnetic moments of localized $f$-electrons hybridize with conduction electrons~\cite{Lacroix1969,Jullien1977,Kaga1988,Read1984,Tsunetsugu1997}. This hybridization creates many-body effects that extend beyond localized atoms, which DFT+$U$, with its focus on local corrections, fails to capture accurately~\cite{Himmetoglu2014}. Additionally, Kondo lattice systems exhibit significant dynamical correlations, with electronic states evolving with temperature~\cite{Doniach1977,Gegenwart2008}. Since DFT+$U$ is a static mean-field approach, it cannot account for these crucial dynamical aspects~\cite{Jones2015}. As a result, the method often predicts incorrect ground state magnetic orders, particularly in materials containing cerium (Ce)~\cite{Matar2013}, ytterbium (Yb)~\cite{Rajabi2024} or uranium (U)~\cite{Belkhiri2019}. These challenges highlight the need for more advanced computational methods to accurately describe the magnetic properties of these complex systems.

The recent development of machine learning assisted discovery has taken route in condensed matter physics, through a combination of systematic search, prediction, and generative algorithms~\cite{Butler2018,Sanchez2018,Stanev2018,Iwasaki2019,Ishii2023}. Machine learning offers a promising alternative to overcome the limitations of traditional computational methods like DFT+$U$. Machine learning techniques can capture complex patterns and interactions that are often missed by conventional approaches~\cite{Zdeborova2017,Carleo2017}. Recent efforts have been made to predict magnetic orders, including for U compounds~\cite{Ghosh2020}, using machine learning models\cite{Katsikas2021,Xia2022,Long2021,Merker2022,Kaba2023,Pant2023,Singh2023}. However, these models typically rely on DFT results as input data. Since accurate DFT calculations are computationally expensive and require extensive tuning, creating a large and reliable dataset is difficult. Therefore, developing machine learning models that can predict magnetic order solely based on structural information is crucial. This approach not only reduces the dependency on labor-intensive DFT calculations but also enables the rapid screening of new materials, accelerating the discovery of compounds with desirable magnetic properties.

%The application to strongly correlated matter is exceeding important, due to the struggle of conventional theoretical methods to predict the ground state properties. 

In this article, we present a machine learning approach to predict the magnetic ground state of U-based materials. We focus on U compounds because they exemplify the complexity of strongly correlated systems. The U 5$f$ bands can manifest as either itinerant or localized, and the crystal electric field (CEF) splitting, spin-orbit coupling (SOC), and Kondo hybridization further complicate ground state calculations~\cite{Smith1983,Endstra1993,Grunzweig1968,Guertin2012,Wu2020,Yuan2021,Sheng1994}. In U-based systems, the interplay of these mechanisms leads to a variety of observed phenomena, including multipolar order~\cite{Mydosh2020}, itinerant ferromagnetism~\cite{Paolasini1996}, heavy fermion behavior~\cite{Si2013}, and topological superconductivity~\cite{Aoki2022}. The intricate and sensitive nature of the U environment makes DFT calculations difficult and often unreliable. On the other hand, our machine learning approach shows very promising results. When applied to an untrained set of known ground states, our random forest classifier achieved a 60.2\% accuracy, compared to a random guess accuracy of 33.3\%. Further limiting our dataset to ternary compounds improves our accuracy to 63.2\%, which rivals previous results with DFT inputs~\cite{Ghosh2020}. 
Given that our input relies solely on structural information and does not require DFT calculations, this is a considerable improvement for the search of novel material properties, including quantum spin liquids, alter-magnetism, and magnetic topological insulators, which have applications for spintronic devices and quantum computing. 

% \vspace{5mm}

\section{Methods}\label{sec:methods}

\begin{figure*}[!ht]
    \centering
    \includegraphics[width=\textwidth]{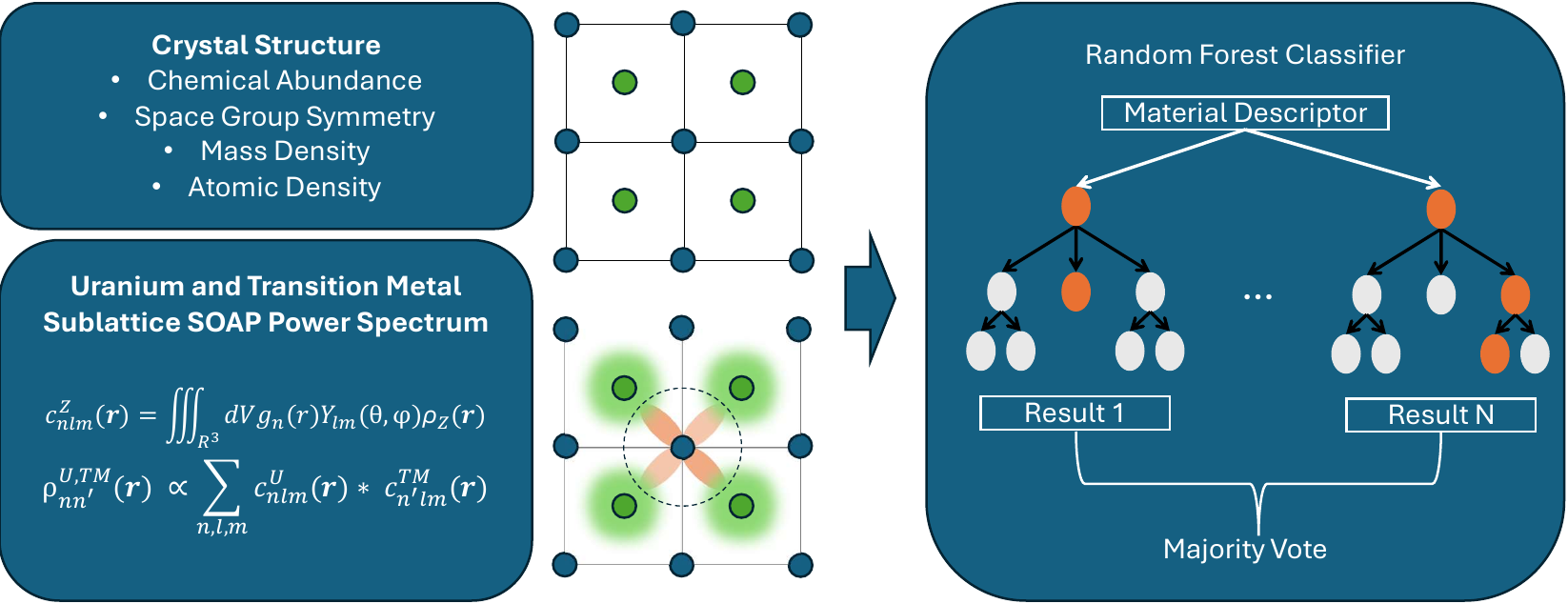}
    \caption{Diagram visualization of the material descriptor to encode the physical properties, and approximated electronic properties with the smooth overlap of atomic positions (SOAP) power spectrum. For the calculation of the SOAP overlap coefficient ($c^{Z}_{nlm}$), $g_n$ is the radial basis function, $Y_{lm}$ is the spherical harmonic, and $\rho_Z$ is the Gaussian smoothed atomic density, following the equation: 
    $\rho_Z(\textbf{r}) = \sum\limits_{i}^{|Z_i|} e^{-1/2 \sigma^2 (\textbf{r}-\textbf{R}_i)^2}$.
    The material descriptor is the input to a random forest classifier, which uses a majority vote to decide the class label and confidence.}
    \label{fig:cartoon}
\end{figure*}

\subsection{Random Forest Classifier} \label{RF}

A random forest~\cite{Breiman2001} (RF) classifier is an ensemble of different decision trees whose classification results are averaged to calculate not only a predicted class for a given object but also a measure of confidence (as seen in Fig.~\ref{fig:cartoon}), usually given by the percentage of decision trees which agree with the chosen class label~\cite{Breiman2001}. A decision tree is a machine learning classification method that has been studied extensively in the literature~\cite{Biau2016}. 
In its simplest form, a decision tree receives a fixed set of variables that describe an object and classifies the object into one of several predefined categories. This classification is based on a series of conditional filters applied to the input variables at each node in the tree.
The nodes at the bottom of a decision tree, known as leaves or leaf nodes, contain final classifications.
In general the larger the distance from the input node to the leaf nodes (the depth $D$), the more accurate the decision tree is as a classifier. However, a very large $D$ is usually indicative of over-fitting to the training data.

In a RF, the different decision trees are randomly tuned with different hyper-parameters such as regularization strength, tree depths, and splitter heuristics, and each tree may only have access to a limited number of input features~\cite{Breiman2001}. Random forests have been shown to have a distinct empirical advantage over single decision trees, with higher explanatory power as well as much less tendency to over-fit to training data~\cite{Biau2016}.

\subsection{Training}
\label{training}

Training a decision tree to best explain a set of known training data samples is a nondeterministic polynomial time (NP) - hard problem, which means it is reducible to a class of computational problems which are believed to not be solvable in polynomial time. Therefore, most of the time a greedy approach is taken to train decision trees~\cite{Hyafil1976}. Starting at the root node, all possible conditions are evaluated and scored by some heuristic, which is usually information gain or a similar criterion~\cite{Quinlan1986}. The routine for choosing the optimal condition at a node is known as the splitter, and it will vary depending on many factors including the type of input features (ie. float, integer, string, etc.), the type of condition desired (ie. classification, regression), and the regularization criteria, among others. The condition with the highest heuristic score should be chosen for the root node. This process is then repeated recursively at each new branch of the tree, creating nodes that operate on the remaining portions of the dataset at each branch. This continues until a node is reached where a single conditional statement can correctly classify all remaining samples in that branch.~\cite{Quinlan1986}. 
In practice, however, performing this training procedure will lead to a very large $D$, and significant over-fitting on the training data. Therefore, a regularization method is critical when training a decision tree to combat over-fitting, and the specific regularization must be tailored to the classification task of interest. Commonly utilized methods of regularization include artificially limiting the $D$, pruning nodes which give minimal information gain over the entire dataset, and requiring a minimum number of training samples to reach each leaf~\cite{Rokach2005}. Often, multiple types of regularization are used in conjunction with one another.

We implement a RF classifier composed of $100$ decision trees using Scikit-Learn~\cite{scikit-learn}, with the Gini impurity~\cite{gini-impurity} as a splitter heuristic and considering a maximum number of $10$ features when looking for the best split at each node. Independent training for each decision tree is achieved through bootstrap sampling and random feature selection~\cite{Breiman2001}. During training, we randomly split the dataset of compounds with known magnetic orders into a training set of $70\%$ of the data-points, and a test set with the remaining $30\%$. The random choice of training data is stratified along the magnetic order so that each order is represented equally in both the training and testing datasets. 

% We train a random forest classifier composed of $100$ decision trees, using the Gini impurity~\cite{gini-impurity} as a splitter heuristic, and considering a maximum number of $10$ features when looking for the best split at each node. Our implementation using Scikit-Learn~\cite{scikit-learn} achieves independent decision tree training through bootstrap sampling and random feature selection~\cite{Breiman2001}. During training, we randomly split the dataset of compounds with known magnetic orders into a training set of $70\%$ of the data-points, and a test set with the remaining $30\%$. The random choice of training data is stratified along the magnetic order so that each order is represented equally in both the training and testing datasets.

\begin{figure}[!b]
    \centering
    \includegraphics[width=\linewidth]{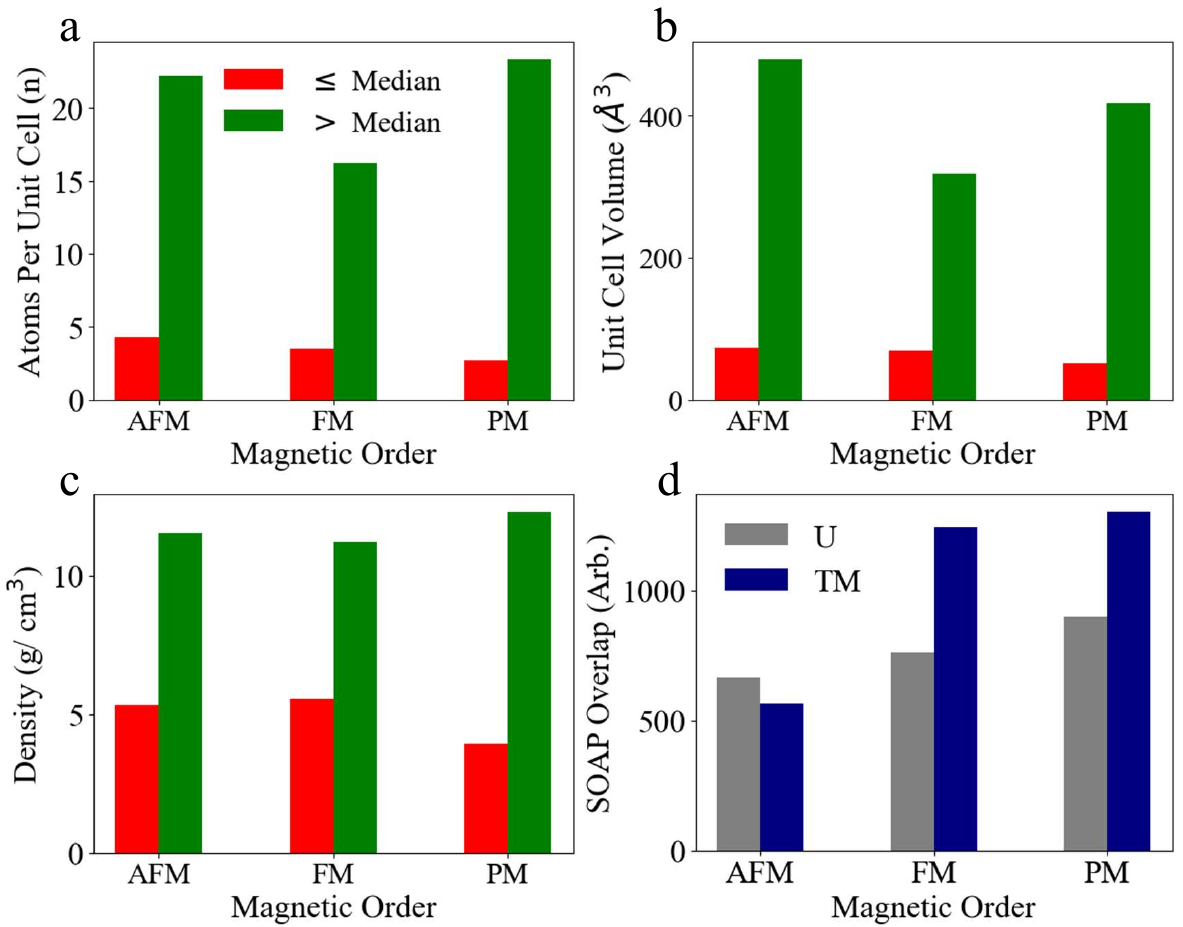}
    \caption{A median-analysis for \textbf{(a)} the number of atoms per unit cell, \textbf{(b)} the unit cell volume, and \textbf{(c)} mass density, grouped by experimental magnetic order. 
    \textbf{(d)} A bar plot comparison of class labels through the mean SOAP overlap of the uranium and transition metal sites, where all transition metal element SOAP overlaps are summed.}
    \label{fig:features}
\end{figure}

\subsection{Data and Descriptor}
\label{descriptor}

We consider a dataset of $1300$ total U compounds, $436$ of which have a known magnetic ordering, which is either anti-ferromagnetic (AFM) or ferromagnetic (FM), or has no magnetic order, which is denoted as paramagnetic (PM). 
The structural details are gathered from the Materials Project\cite{materialsproject} database, which includes a subset of roughly $800$ materials. Using this information, each compound is described by $163$ features in total. $120$ of these features are simply the relative abundances of each element in the compound, so this vector is very sparsely non-zero. The remaining features, as seen in Figure~\ref{fig:cartoon}, include the space group, atoms per unit cell, unit cell volume, mass density, and the smooth overlap of atomic positions (SOAP) power spectrum. 

The SOAP algorithm~\cite{De2016,Jager2018,Bartok2013,Willatt2018} provides a comprehensive structural descriptor that captures the local atomic environments.
Each atomic site is modeled by spherical harmonics and radial basis functions, and the overlap is computed through a real-space Gaussian approximation~(see Fig.~\ref{fig:cartoon}). We implement the SOAP algorithm, using the DScribe python package~\cite{dscribe2,DScribeSOAPdoc}, to account for the U 5$f$ electrons through the parameters: $n_\mathrm{max}$ = 5, 
$l_\mathrm{max}$ = 3 
and $r_\mathrm{cutoff}$ = 
5~\AA. To further simplify the SOAP power spectrum, we limit the calculation to the U and transition metal (TM) sub-lattice, and we record the sum for all quantum numbers and sites for a given atomic species. For example the U overlap (U$_\mathrm{overlap}$) of a U-Fe sub-lattice would be computed as:
\begin{equation}
    \label{eqn1:overlap}
    \mathrm{U}_\mathrm{overlap} = \sum_{i}~(~ \sum\limits_{n'=1}^{5} \rho_{nn'}^{\mathrm{U}_i,\mathrm{U}}(\textbf{r})+ \sum\limits_{n'=1}^{5}\rho_{nn'}^{\mathrm{U}_i,\mathrm{Fe}}(\textbf{r})~)
\end{equation}
While summing over all sites and quantum numbers loses some information, it is necessary to compare diverse sets of materials. 

A statistical representation of our known dataset is shown through a median analysis of our material descriptor, grouped by experimental magnetic order~(Fig.~\ref{fig:features}). The unit cell volume and number of atomic sites appear to have similar distributions~(Fig.~\ref{fig:features}a,b), which is reflected in the mass density comparison in Figure~\ref{fig:features}c.
The SOAP power spectrum comparison provides more valuable insights. The AFM label is quantitatively different from either the FM or PM label, where the U$_\mathrm{overlap}$ is larger than the TM$_\mathrm{overlap}$ (Fig.~\ref{fig:features}d). This suggests that U$_\mathrm{overlap}$ and TM$_\mathrm{overlap}$ may play an important role in determining the magnetic order.

\begin{figure*}[!ht]
    \centering \includegraphics[width=\textwidth]{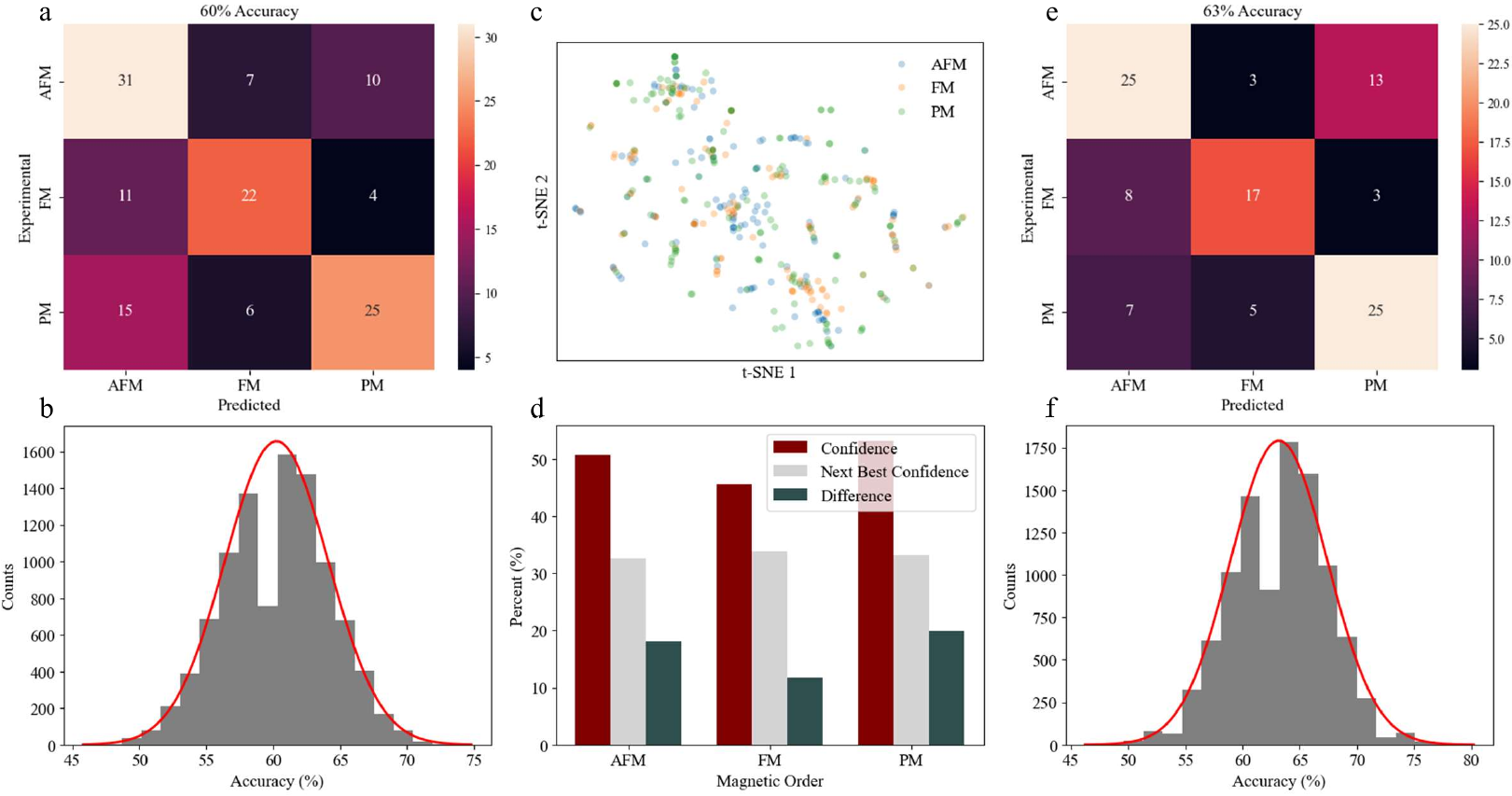}
    \caption{\textbf{(a)} The confusion matrix presentation of a random forest for a training set with 60\% accuracy. \textbf{(b)} A histogram of 10,000 stochastic training splits, fitting to a Gaussian distribution described by $\mu=60.2\%$ and $\sigma=3.8\%$.
     \textbf{(c)} The t-SNE visualization for our material descriptor set.
     \textbf{(d)} The average confidence, next best confidence and difference for the predictions of unknown magnetic orders.\textbf{(e)} A sample confusion matrix, limited to ternary compounds, and the histogram for 10,000 training splits, giving $\mu=63.2\%$ and $\sigma=4.2\%$. }
    \label{fig:results}
\end{figure*}

\section{Results} \label{sec:performance}

\subsection{Validation}
\label{valid}

Figure \ref{fig:results}a shows a confusion matrix of the magnetic order predicted by our RF classifier versus the true labels for the test dataset. The confusion matrix allows us to compare the predictions for each magnetic order. The FM category has a slightly lower accuracy at $56.8\%$, while the AFM and PM labels score at $70.8\%$ and $65.2\%$, respectively. From this, it can be seen that our model does not suffer from any form of mode collapse or over-fitting, and we expect the prediction accuracy to grow as we acquire more training data. To acquire the true accuracy, we display a histogram of 10,000 stochastic training splits, which fits to a normal distribution with a mean ($\mu$) of $60.2\%$ and a standard deviation ($\sigma$) of $3.8\%$~(Fig.~\ref{fig:results}b). With three possible class labels, we expect a minimum of $33.3\%$ accuracy by random chance. In that view, our result is as good as we might expect for our very limited dataset size.

We perform a t-distributed stochastic neighbor embedding (t-SNE)~\cite{t-SNE} visualization to qualitatively capture the dependence of magnetic order on our descriptor~(Fig~\ref{fig:results}c). We can see that the compounds do not appear to be most easily grouped by magnetic order, implying that linear methods of analysis, such as support vector decomposition or principal component analysis, will not be enough to accurately predict magnetic order in new U compounds.

For a comparison with a DFT-assisted RF classification~\cite{Ghosh2020}, we limit our dataset to 351 U ternary compounds. Figure~\ref{fig:results}e presents a confusion matrix from our RF classification, where the true accuracy is $63.2\%$~(Fig.~\ref{fig:results}f). This is comparable to a previously reported RF classifier~\cite{Ghosh2020}, which used DFT-calculated features to model 136 U ternary compounds. 
While the DFT-suported descriptor yields a slightly higher accuracy at $68.9\%$~\cite{Ghosh2020}, our model does not rely on assumptions of electron dynamics, which would be counter-intuitive for predicting strongly correlated materials. Furthermore, Ghosh et al.~\cite{Ghosh2020} reported a 15\% improvement in classification accuracy by incorporating an Orbital Field matrix (OFM) into the material descriptor. As a similar algorithm to the SOAP power spectrum, the OFM effectively encoded the crystal structure, valence occupation, and local chemical environments~\cite{Ghosh2020},  further supporting the local environment being fundamental to classification accuracy. 

% The advantage of the SOAP algorithm provides similar information on the atomic spacing, coordination number and electron wave-function symmetry. Additionally,  
\subsection{Prediction}
\label{predict}

We subsequently use the trained random forest classifier, as seen in Figure~\ref{fig:results}a, to predict the magnetic order for $376$ U compounds.  Nearly half of these materials are predicted to be AFM (185), while the other half is split between FM (46) and PM (136). The average confidence for each category is near $50\%$, as seen in Figure~\ref{fig:results}d, with the FM category being the lowest at 46\%. The next best confidence (NBC) is also plotted for a quantitative representation of the uncertainty. If a large population of the random forest is uncertain, the NBC will also be large. The FM label has the largest NBC, while the AFM and PM labels sit just above $30\%$. Since the confidence for the FM label is slightly smaller, it manifests as a comparatively smaller difference, as seen in Figure~\ref{fig:results}d. The smaller confidence is likely related to the FM materials taking up a smaller part of our known dataset, resulting is smaller $D$ for FM decisions.

% From our highest confidence predictions, Th$_3$U and U$_2$TeSe both have a confidence of 74\% for AFM labeling and next-best confidence difference of 59.5\% and 57\%, respectively.

To compare our prediction with an industry standard, we use the the Materials Project~\cite{materialsproject} (MP) magnetism simulation as a control. For the majority of unknown materials, MP uses a numerical simulation to calculate the lowest energy, collinear magnetic configuration~\cite{MatProjMagMeth}. The simulation begins with FM ordering, and progresses to either a ferrimagnetic (FiM), AFM, or non-magnetic (NM) --- synonymous with PM~\cite{MatProjMagMeth}. As we compared our predictions to MP, we realized that none of the compounds had a simulations result of AFM. To console this, we consider two comparisons in Table~\ref{table:prediction}: (1) considering FiM as AFM, FM as FM, and NM as PM, and (2) considering FM/FiM as magnetic and AFM/PM/NM as non-magnetic. In the first comparison,the FM category has the best agreement, at ~69.6\%, while the AFM and PM predictions are far lower at 14.6 and 25.7\%, respectively. For the second comparison, our FM predictions have a 89.1\% agreement with MP simulations, while our AFM and PM predictions have an 18.7\% alignment with NM simulations. The FM bias has been noted in other machine learning papers~\cite{Long2021,Merker2022,Kaba2023} and the MP documentation~\cite{MatProjMagMeth}, and additional effort have been performed with high throughput predictions using DFT methods~\cite{Horton2019,Frey2020}.

\begin{table}[!t]
\centering
\caption{Prediction alignment with Materials Project.}
\begin{tabular}{l@{\hskip 10pt} c@{\hskip 10pt} c@{\hskip 10pt} c@{\hskip 20pt} c@{\hskip 10pt} c}
 \toprule
 Magnetic& \multicolumn{3}{c}{Comparison} & \\

Order & & (1)  & (2) \\  
 \midrule
 AFM   & & 14.6\% &  - & \\ 
 FM    & & 69.6\% &  89.1\% & \\
 PM    & & 25.7\% &  18.7\% & \\   
 \bottomrule
\end{tabular}
\label{table:prediction}
\end{table}

% This explains the misalignment for AFM and PM labels, while our FM showed a promising comparison with MP simulations Table~\ref{table:prediction}.

% When comparing for net-magnetism, we find excellent agreement for a non-zero net-magnetism, while the zero net-magnetic is below 20\%. This suggests that 10\% of our FM labels, may rely on higher order interactions; and, many of these net-zero magnetism compounds, may have their magnetic order suppressed by long rangeinteractions. 

\section{Discussion} \label{sec:discussion}

Our RF classification exhibits reliable classification performance. The standard deviation of our true accuracy is small~(ie. $\sigma / \mu = 0.063$), indicating that our training sets provide consistent representation of our material dataset. Additionally, there is minimal preference towards a specific class label seen in our validation accuracy~(Fig.~\ref{fig:results}a,e) and prediction confidence~(Fig.~\ref{fig:results}d). This is notable given the small size of our dataset. While many classification methods exist, our choice for a RF approach is supported by our t-SNE visualization (Fig.~\ref{fig:results}d).
This illustrates that the correlation between our structural descriptor and magnetic order cannot be captured by a linear method of analysis. 
Alternatively, a neural network would risk over-fitting due to the high number of parameters relative to data points.
The RF algorithm is ideal for small datasets such as our own, due to the regularization of hyper-parameters between each decision tree (see Sec.~\ref{RF}).
This provides protection from over-training, which is a well-known risk in neural networks. 

The MP simulations provide an interesting comparison for our prediction results, where the unaligned predictions may have magnetic order emerge or diminish due to higher-order interactions. The MP simulations are a scalar, Ising-type model, which only captures on-site correlations between atomic moments and have been reported to have bias towards the FM order~\cite{Kaba2023}. While a spin-lattice description may be a reasonable model for the FM case, which has excellent agreement with MP simulations~(Table~\ref{table:prediction}), the AFM and PM ordering requires a more detailed model. The Kondo lattice description fosters a competitive relationship between Kondo resonance spin-scattering and magnetic ordering from AFM-type Ruderman-Kittel-Kasuya-Yosida (RKKY) exchange interactions. Since these interactions rely on itinerant electron behavior~\cite{Tsunetsugu1997,Lacroix1969,Doniach1977,Gegenwart2008}, on-site correlations will not be enough to predict the magnetic ground state. 

The limiting factor for our machine learning prediction is how well our material descriptor represents the underlying phenomenon. We chose the SOAP power spectrum~\cite{DScribeSOAPdoc}; however, alternative algorithms, such as the Coulomb and orbital field matrices, provide similar information. The SOAP algorithm is advantageous in that it uses a real-space Gaussian density approximation with spherical harmonics~\cite{DScribeSOAPdoc}, which is absent in the Coloumb matrix, and the OFM requires information on the valence occupations, which must be estimated or calculated from numerical methods. The success of the OFM in Ghosh et al.~\cite{Ghosh2020} reaffirms the importance of local symmetry information. Further, our material descriptor can still be tuned. The SOAP overlap algorithm~(Sec.~\ref{descriptor}, Eqn.~\ref{eqn1:overlap}) can be tailored to exclusively record the U 5$f$ overlaps. It may also be productive to expand our feature list to include the local atomic environment through the Wyckoff letter. Alternatively, we could condense the $120$ chemical abundance elements, reducing the total length and increasing the information density. With an already substantial accuracy, Bayesian optimization~\cite{garnett_bayesoptbook_2023} can further help optimize the chosen features and hyper-parameters for our RF classification.

% The Kondo lattice is known for a competitive relationship between , both of which are mediated by conduction electrons. This dynamic and many-body effect fosters a complex expression for U 5$f$ electrons, which goes beyond on-site correlations, and may explain the discrepancy for the AFM and PM categories.

U-based materials offer a platform for discovering new quantum materials. Our descriptor shows the ability to intuit the magnetic ground state with reliable accuracy, despite the complex nature of the underlying mechanisms. The U-based Kondo systems are more complex that the Ce and Yb counterparts, due to a more extended wave-function and multiple $f$ electrons in the U$^{4+}$ valence state. This allows for a stronger hybridization effects, due to the extended nature~\cite{Smith1983,Endstra1993}, and the potential for an orbital-selective Kondo lattice, where magnetic order will coexist with Kondo coherence~\cite{Giannakis2019,Siddiquee2023,Lin2024}. The ground state calculation must incorporate the many-body correlations between conduction electrons and the localized 5$f$ moments~\cite{Matar2013,Rajabi2024,Belkhiri2019}. This usually requires computationally expensive calculations, such as dynamical mean-field theory (DMFT). Instead, we address these factors with the SOAP algorithm to encode the local symmetry information and wave-function overlap, providing crucial information to predict the magnetic expression of U 5$f$ local moments with machine learning.

% The magnetic character of the U 5$f$ electrons are more intricate due to the natural presence at the Fermi energy, which fosters strong hybridization effects. Further, the uranium valence commonly possesses multiple 5$f$ electrons. This can lead to the occurrence of an orbital selective Kondo lattice, where magnetic order coexists with Kondo coherence.  Instead, we have treated our curated dataset of uranium compounds with a structural descriptor and random forest classification for a computationally inexpensive guide for material discovery.

% To validate our predictive capacity, we synthesize the compounds Th$_3$U and U$_2$TeSe, which both have a high confidence for AFM ordering. \textcolor{red}{add experimental verification here to support predictions}

\section{Conclusion}
Our study introduces a fundamental structural descriptor that effectively guides material discovery, particularly in cases where standard numerical methods fall short in capturing underlying phenomena. The implementation of a RF classifier has demonstrated a powerful and computationally efficient predictive capability. By integrating symmetry considerations with magnetic order predictions, this work significantly aids in the discovery of new quantum materials.

\begin{acknowledgments}
We acknowledge fruitful discussions with Zohar Nussinov. The work at Washington University is supported by the
National Science Foundation (NSF) Division of Materials Research Award DMR-2236528. C. Broyles
acknowledges the NRT LinQ, supported by the NSF under Grant No. 2152221.
\end{acknowledgments}

\appendix

\section{Software information}
The following software packages were used in the implementation of this code: 
(1) The random forest classifier and Gini-impurity from Scikit-learn~\cite{scikit-learn}, (2) t-SNE from Seaborn~\cite{Waskom2021} and  (3) the Smooth Overlap of Atomic Positions from DScribe~\cite{dscribe2}. The crystallographic information file (CIF), mass density, and atomic density were collected using the Materials Project\cite{materialsproject} application programming interface (API).

% \bibliography{main}
% \bibliographystyle{aasjournal}

\end{document}